# Ballistic-diffusive Phonon Heat Transport across Grain Boundaries

Xiang Chen[1*], Weixuan Li[1], Liming Xiong[2], Yang Li[1], Shengfeng Yang[1], Zexi Zheng[1], David L. McDowell[3,4], and Youping Chen[1]

[1]Department of Mechanical and Aerospace Engineering, University of Florida, Gainesville, FL 32611, USA

[2]Department of Aerospace Engineering, Iowa State University, Ames, IA 50011, USA

[3]Woodruff School of Mechanical Engineering, Georgia Institute of Technology, Atlanta, GA 30332, USA

[4]School of Materials Science and Engineering, Georgia Institute of Technology, Atlanta, GA 30332, USA

**ABSTRACT:** The propagation of a heat pulse in a single crystal and across grain boundaries (GBs) is simulated using a concurrent atomistic-continuum method furnished with a coherent phonon pulse model. With a heat pulse constructed based on a Bose-Einstein distribution of phonons, this work has reproduced the phenomenon of phonon focusing in single and polycrystalline materials. Simulation results provide visual evidence that the propagation of a heat pulse in crystalline solids with or without GBs is partially ballistic and partially diffusive, i.e., there is a co-existence of ballistic and diffusive thermal transport, with the long-wavelength phonons traveling ballistically while the short-wavelength phonons scatter with each other and travel diffusively. To gain a quantitative understanding of GB thermal resistance, the kinetic energy transmitted across GBs is monitored on the fly and the time-dependent energy transmission for each specimen is measured; the contributions of coherent and incoherent phonon transport to the energy transmission are estimated. Simulation results reveal that the presence of GBs modifies the nature of thermal transport, with the coherent long-wavelength phonons dominating the heat conduction in materials with GBs. In addition, it is found the phonon-GB interaction can result in the reconstruction of the GBs.

## 1. Introduction

Experimental studies of phonon transport using direct photoexcitation have led to observations of many new phenomena. One observation that is of principal significance is the non-diffusive heat propagation in non-metallic crystals, especially those with interfaces [1]. The ballistic-diffusive phonon transport and its interaction with interfaces can be manipulated to control the thermal conductivity of materials, which can lead to significant scientific and technological advances. The need for a better understanding of phonon transport has in turn driven the development of experimental techniques such as the time domain thermo-reflectance (TDTR) and related ultrafast pump-probe techniques that utilize ultrafast laser pulses. The extremely fine temporal resolution associated with the ultra-short laser pulses offers a desired fine spatial resolution [2], enables the precise measurement of interfacial thermal conductance [3-7], and has provided useful insights into the nature of phonon transport across materials interfaces. For example, one recent experiment has reported that the measured thermal conductivity in superlattices by TDTR grows linearly with the increasing total thickness of the superlattice, which has been interpreted as an experimental evidence for coherent phonon transport in semiconductor superlattices [8]. Another more recent experimental study has observed a minimum in the lattice thermal conductivity in epitaxial oxide superlattices as a function of interface density, and this minimum has been believed to be a crossover from incoherent to coherent phonon scattering [9]. Computational approaches to advance the understanding of phonon thermal transport in materials include the lattice dynamics (LD) method, the atomistic Green's functions (AGF), Boltzmann transport equation (BTE) approaches, and the classical molecular dynamics (MD) method. The AGF approach has emerged as a useful technique to calculate the interaction of phonons with extended structure, which can be used to find energy transmission across an interface and the interfacial thermal conductance. However, like the LD method, AGF approach is a frequency domain method, it typically is applied to a periodic system, and is restricted to coherent scattering and systems with nanoscale structural feature [10]. BTE method is powerful for the study of diffusive heat conduction, therefore it neglects wave effects such as wave interference and phonon focusing [11]. To understand phonon transport across materials interfaces, MD is essentially the only simulation method currently available that does not need assumptions on the nature of phonon transport or phonon-interface scattering mechanisms [11, 12], e.g., whether phonons are coherent or incoherent, to be explicitly incorporated into the equations that govern the simulation, all phenomena emerge naturally in the simulation as the consequence of the laws of classical mechanics and the interatomic potential that describe the interaction between atoms. The major limitations of the classical MD for the study of phonon-interface interaction are that it can only capture classical effects, the available interatomic potentials are not accurate for quantitative predictions of thermal conductivity, and the maximum phonon wavelength that can be present in the simulation is only the length of the MD simulation cell. There is a gap in length scale between the simulated and actual structure, which cannot be bridged by using periodic boundary conditions [13]. In order to gain a quantitative understanding of the role of long-wavelength phonons and their interaction with grain boundaries, in this work we use a coarse-grained atomistic method, the concurrent atomistic-continuum (CAC) method, to simulate the transient propagation of heat pulses in single and polycrystalline materials. It should be noted that compared with coherent interfaces in superlattices, phonon transport across incoherent interfaces such as high angle grain boundaries are less studied and are more challenging to experimental and other computational approaches due to the microstructural complexity. As pointed out by Chen [14], the theoretical solution of the non-diffusive heat conduction in materials with complicated geometrical configurations is a difficult research topic.

The CAC method is based on a unified atomistic-continuum formulation of conservation laws as an extension of Kirkwood's statistical mechanical theory for transport processes [15, 16]. It differs from the Kirkwood's formulation by a concurrent two-level description of a crystalline material as a continuous collection of material points (unit cells) while each material point possesses internal degrees of freedom that describe the movement of atoms inside each unit cell. This two-level structural description of crystalline materials leads to a concurrent atomistic and continuum method for multiscale simulation of transport processes within one single theoretical framework [17-19]. As a result, CAC can simulate complex crystalline materials, reproduce both acoustic and optical phonon branches, resolve details to full atomistic resolution at interfaces or in other regions of interest while coarse-graining to thousands of atoms per element elsewhere, and reflect dynamics on the length and time scales of interest. The CAC method has been demonstrated to be able to predict the dynamics of defects including dislocations [20-26] and cracks [27-29], the dynamics of phase transformations [30], and to reproduce the full sets of phonon branches of polyatomic materials in both the atomic and the coarse-grained regions [31]. Specifically, our simulation results show that CAC can reproduce the exact dynamics of an atomistic system if modeled with the finest mesh, and if discretized with finite elements, it can precisely predict the dynamics of phonons whose wavelengths are longer than a critical wave length determined by the size of the element, and hence can be used to investigate the dynamics of long-wavelength phonons. In this study, the CAC method is employed, for the first time, to simulate a heat pulse experiment. The objective of this work is to simulate and to visualize the transient propagation of a heat pulse in both single crystal and across grain boundaries using CAC. We employ a phonon representation of heat pulses composed of spatiotemporal Gaussian wave packets to mimic the coherent excitation of a nonequilibrium phonon population by ultrashort laser techniques. Note that a coherent wave/transport is usually defined to be the one that has constant phase, or that preserves the phase of the wave, whereas incoherent wave/transport has a phase that fluctuates at random [32].

The rest of this article is organized as follows. The methodology is presented in Sec. 2, including the phonon representation of heat pulses and CAC computer models; for verification and validation purposes, the phonon dispersion relation calculated from CAC simulation results are compared with that from analytical lattice dynamics calculation; the effect of the numerical interfaces between the atomistic and continuum descriptions is quantified through the comparison between the wave propagation in single crystals with uniform and non-uniform meshes. In Sec.3, we present the CAC simulation results of phonon pulses propagation in three samples with different number of GBs; the wavelength-dependent energy transmissions are quantified through the simulation of monochromatic phonon wave-GB interaction, and the phonon induced GB structural change is presented. This paper is then ended with a brief summary and discussions in Sec.4.

## 2. Methodology and Verification
### 2.1 The phonon representation of ultrafast heat pulses and its application in CAC

There is currently no well-established methodology of using MD to simulate transient propagation of heat pulses. Based on the fluctuation-dissipation theorem, the Green-Kubo method [33, 34] can be used to compute the thermal conductivity of materials, however, only for homogeneous systems under thermal equilibrium. The nonequilibrium MD, or the direct method, mimics the heat source and sink in experiment, and then uses Fourier's law to quantify the thermal conductivity, nonetheless, only applies to diffusive heat conduction at steady-state [35-37]. In the past decades, a few attempts have been made to study the transient heat pulse propagation using MD thermostats to model the heat pulses [38-41]. However, thermostat algorithms thermalize phonons and tend to destroy phonon coherency [42], and hence are not suited for the simulation of the coherent phonon excitation by ultrafast laser pulse experiments. To mimic the response of materials to ultrafast laser pulses, we have developed a phonon representation of heat pulses, termed a coherent phonon pulse (CPP) model, and have demonstrated the heat pulse model through quantifying the Kapitza resistance of GBs to heat flow and identifying the phonon-GB scattering mechanism in a nanoscale bi-crystalline silicon [41]. It is worth noting that the laser-induced phonon generation is still a subject of debate, especially regarding the frequency and spatial distribution [43-47]. In experiments, there are many factors that determine the phonon generation by a laser pulse, including the laser intensity, wavelength of the laser pulse, time duration, spot size, background temperature, whether there is a presence of a metal film, etc. As pointed out by Wolfe [1], because phonon generation depends on the particular electron-phonon interactions, and because high-frequency phonons can anharmonically decay into lower frequency phonons, there is no generic formula determining the frequency distribution of the nonequilibrium phonons that are generated. Nevertheless, the number of phonons excited can be closely related to the concept of temperature, and an ideal heat pulse is the one that contains a Bose-Einstein (Planckian) distribution of phonons [48], in which the phonon number n indexed by the wavevector $\mathbf{k}$ and the phonon branch $\nu$ is related to the temperature T by

$$n(\mathbf{k}, \nu) = [\exp\left(\frac{\hbar\omega(\mathbf{k},\nu)}{k_BT}\right) - 1]^{-1} \qquad (1)$$

where ω is the phonon frequency, ℏ the reduced Planck constant and $k_B$ the Boltzmann's constant. Then, according to [49], the phonon number of a specific mode is related to the displacement of the given mode.

Since the CAC method is a coarse-grained method derived bottom up from the atomistics, and can adapt the viewpoint of phonons to describe heat and temperature in terms of mechanical vibrations, the atomic-vibration based CPP model naturally fits in the framework of CAC. In addition, the theoretical foundation determines that CAC can be used to study highly nonequilibrium dynamic processes of the phonon propagation without the need for additional assumptions of rules or preconceived understanding of the underlying mechanisms, the same as MD. Therefore, in this work, to simulate the dynamic process of an ultrafast heat pulse propagating in a solid material by CAC, we construct a radially distributed phonon heat pulse [41] based on experimental observations of radial phonon distributions in materials induced by the photoexcitation [50]. The heat pulse is then applied to a local heating area in a CAC computer model, containing phonons with all wave vectors and frequencies that the computer model allows. The propagation of the heat pulse and the interactions between phonons and grain boundaries, surface boundaries, as well as phonon-phonon interaction, then naturally emerge in the CAC simulation as a consequence of the governing equations and the interatomic potential.

## 2.2 CAC models of single and polycrystals

The primary objective of this work is to investigate the transient propagation of a heat pulse across GBs. For the purpose of comparison, the propagations of heat pulses in single crystals are also simulated. In Fig. 1(a), we present the 2D single crystal model discretized with a uniform quadrilateral shaped finite element (FE), with the edge length of the model being L = 0.5 μm. Each specimen contains 60,000 elements, and each element represents 64 atoms. A four-neighbor Lennard-Jones (LJ) interatomic potential [51] is employed for the description of the force field in the CAC simulations, yielding the crystal structure of (111) plane in FCC copper. Note that the CAC computer models of polycrystals necessarily contain atomistically resolved fine mesh for the GB regions and coarsely meshed finite elements for regions away from the GBs. Such non-uniform mesh may result in spurious wave scattering by the interface between different numerical resolutions. For verification purpose, the single crystal model is also discretized with a non-uniform mesh, i.e., with both coarse mesh and atomistically-resolved fine mesh, as shown in Fig. 1(b).

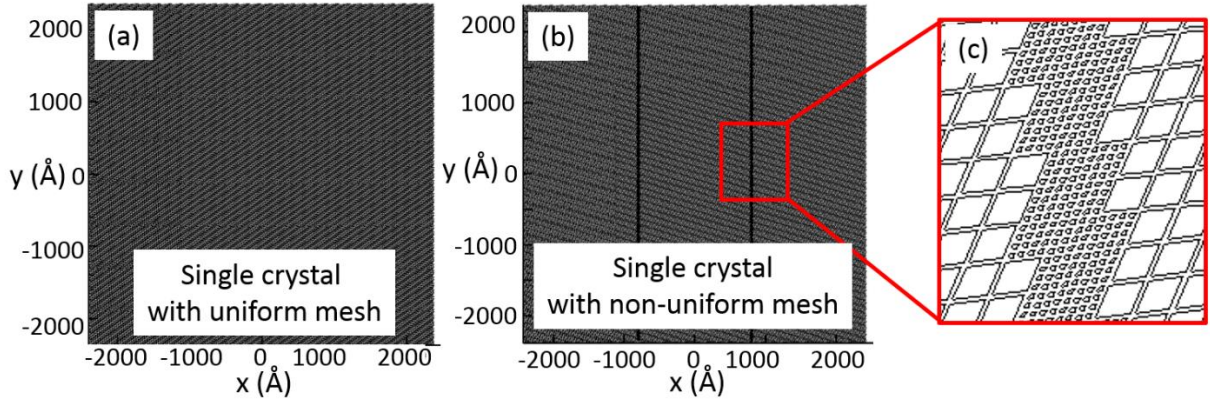

**Fig. 1** Single crystal material sample modeled (a) with purely coarse finite element mesh and (b) with both coarse mesh and fine mesh. (c) is the zoom-in detail of the interface between coarse and fine mesh in (b).

In Fig. 2 we present the CAC models of polycrystals with different number of GBs. Same with the single crystal model in Fig. 1, the edge length of each model is L = 0.5 μm. Considering the symmetry of the structure, the two models displayed in Fig. 2(a) and Fig. 2(b) are named GB-1 and GB-2, respectively. The former contains one GB on each side and the latter two GBs on each side. The GBs in both models have the same atomic structure. The finite element mesh near the GB region is shown in Fig. 2(c), with the region near the GBs being discretized into finite elements with atomic resolution and the region away from GBs being discretized into coarsely meshed finite elements. The orientation of the elements is dependent on the crystallographic direction of the grains, which is determined by the inclination angle of the GB. The details of the atomic arrangements in the vicinity of the GBs are presented in Fig. 2(d). The inclination angle between the crystalline grains is 13.2°. It can be seen from Fig. 2(d) that the GBs are composed of periodically repeated pentagonal structures. In order to compare the phonon transport in single crystals and polycrystals, the orientation of the elements (determined by the crystallographic orientation) in single crystals and that in the heat-source-grain of the polycrystals are the same. A same heat pulse is then applied at the center of all the models to make sure the initialization of the thermal loading is identical in different models. In order to isolate the effect of phonon-GB scattering and to make the condition less complicated, the simulations start at 0 K. However, the phonon-phonon scattering, phonon-defect scattering, as well as the local temperature generated by the heat pulse and the scattering, are all naturally allowed and the entire system is evolved with central difference time integration.

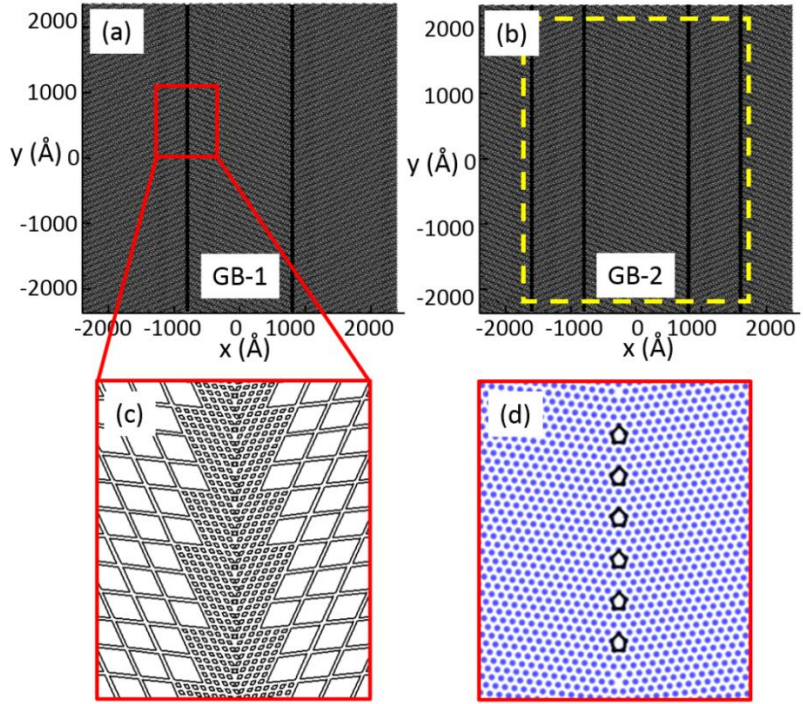

**Fig. 2** CAC models of the polycrystals (a) with one GB (named GB-1) and (b) with two GBs (named GB-2) on each side; (c) finite element mesh near the GB region; (d) the atomic configuration of the GB obtained by mapping from the finite element mesh showing the periodic pentagonal structures in the GBs; The yellow dash-line box in (c) marks the region where the kinetic energy is monitored on the fly to quantify the energy transmission in Section 3.1.

## 2.3 Verification of the CAC method

### 2.3.1 1D phonon dispersion relation

The dynamics of a crystalline material system can be characterized by phonon dispersion relations. The ability to provide a direct test of the correctness of a theoretical method has been considered to be one of the significances of the experimental measurement of the phonon dispersion relation [52]. To test the CAC method and the simulation tool in reproducing phonon dispersion relations, we construct a one-dimensional CAC model of a monatomic chain system that contains 500 linear finite elements in the middle and 1500 atoms at each of the two ends of the specimen, with periodic boundary conditions applied. The schematic sketch of the computer model is presented in Fig. 3 (c). The size of the finite elements in the CG (coarse-grained) region is six times the lattice constant. A four-neighbor Lennard-Jones (LJ) interatomic potential [51] is employed for the description of the force field in the CAC simulations, yielding a lattice constant 2.556 Å as that of the [110] direction in FCC solids.

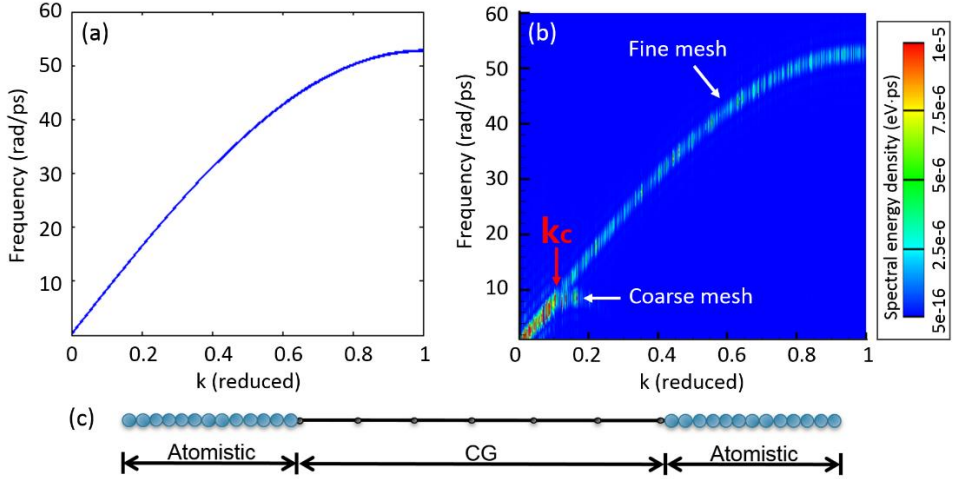

**Fig. 3** (a) Lattice dynamics solution of phonon dispersion relation of a monatomic chain that models the [110] direction of FCC crystal, (b) corresponding phonon spectral energy density calculated from the CAC simulation results, (c) schematic sketch of the CAC model of a 1D atomic chain.

Fig. 3(a) presents the phonon dispersion relation of the monatomic chain analytically calculated using lattice dynamics (LD). Fig. 3 (b) presents the phonon spectral energy density obtained from CAC simulation of the same monatomic system, but modeled with both atomistically resolved fine mesh and coarse FE mesh. The phonon spectral energy density $\phi(\boldsymbol{k}, \omega)$, defined as the averaged kinetic energy per unit cell as a function of wave vector and frequency, can be calculated through computing the velocity-velocity autocorrelation function in MD simulations following the expression [53]

$$\phi(\boldsymbol{k}, \omega) = \frac{1}{4\pi\tau_0 N_T} \sum_\alpha \sum_b^B m_b \left| \int_0^{\tau_0} \sum_{n_{x,y,z}}^{N_T} \dot{u}_\alpha \binom{n_{x,y,z}}{b}; t \right) \times \exp\left[i\boldsymbol{k} \cdot \boldsymbol{r} \binom{n_{x,y,z}}{0} - i\omega t\right] dt \Bigg|^2. \tag{2}$$

in which, $\tau_0$ is the phonon relaxation time, $N_T$ the total number of unit cell, $\dot{u}_\alpha \binom{n_{x,y,z}}{b}; t$ the velocity in the direction $\alpha$ of atom $b$ (with mass $m_b$) inside the unit cell $n_{x,y,z}$, $\boldsymbol{r} \binom{n_{x,y,z}}{0}$ the equilibrium position of each unit cell. Similarly, in CAC simulations, the spectral energy density is computed via post-processing the velocities of atoms in the atomistic region and the FE nodal velocities in the finite element region. In Fig. 3(b), we can clearly identify the phonon dispersion curves from the spectrum. On one hand, we see that the phonon dispersion relations obtained for the atomically resolved regions in the CAC model compare very well to those from LD calculations; this indicates that the CAC method can reproduce the dynamics of atoms in an atomistic system accurately if modeled with the finest FE mesh. On the other hand, it is seen that the phonon dispersion relation of the coarsely meshed FE region and that of the atomistically resolved region overlap with each other only for wavevector smaller than a certain value. This is because coarse-graining cuts off the small wavelength phonons with wave vector larger than $k_c$ or with wavelength smaller than $2\pi/k_c$, in which the critical wavevector $k_c$ with an allowed error $\varepsilon$ is determined by the finite element size ($h = na$) in the CG region according to Eq. (3), in which, $a$ is the lattice constant,

$$k_c = \max\left\{ k : \left| \sin\left(\frac{ka}{2}\right) - \sin\left(\frac{kh}{2}\right) \right| \le \varepsilon \right\} \tag{3}$$

This means that only phonons whose wavelengths are larger than $2\pi/k_c$ can propagate in the CAC model. Nevertheless, since the difference between the phonon dispersion relation for the atomistic and that for coarsely meshed FE region is negligible for $k < k_c$, as shown in Fig. 3(b), it guarantees that a CAC model discretized with coarsely meshed finite elements can precisely predict the dynamics of phonons whose wavelengths are longer than $2\pi/k_c$ and hence can be used as a complimentary tool to MD in predicting the dynamics of long wave length phonons. The critical wavelength is computed to be ~5nm for the material samples considered in this work.

### 2.3.2 2D phonon dispersion relations

Since the computer models used to study the phonon-GB interaction in this work are two-dimensional, we also calculate the phonon dispersion relations for the 2D lattices. Considering the computer model shown in Fig. 1(a) as a part of the bulk crystal, the phonon dispersion relations can be obtained by solving the dynamical matrix of the CG lattices. In Fig. 4, the obtained phonon dispersion relations along [110] direction of the FCC crystal are presented and the CAC results are com-pared to the LD solution, each containing longitudinal acoustic (LA) and transverse acoustic (TA) branches. It is seen from Fig. 4 that, within the range of the allowable wavelength (5nm~250nm) in the simulation model, the phonon dynamics predicted by CAC matches the LD solution well.

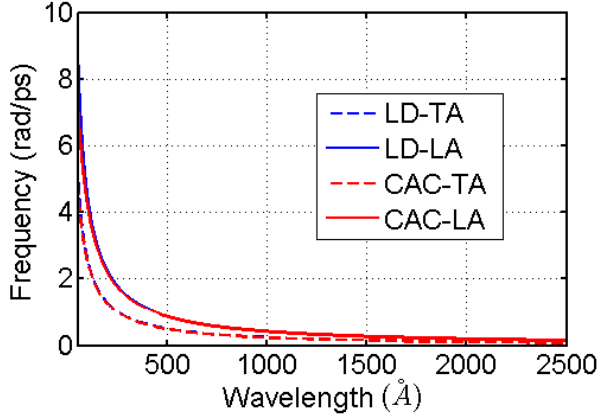

**Fig. 4** Phonon dispersion relations along the [110] direction within the (111) plane of FCC single crystal. The solutions calculated for the coarse-grained lattices in CAC (in red) are compared to the lattice dynamics solution (in blue). The solid lines are LA braches and the dashed lines are TA branches.

### 2.3.3    Continuous waves propagation across atomistic-continuum interface in single crystals

In a concurrent multiscale simulation that contains both continuum and atomistic descriptions, the spurious wave reflection at the atomistic-continuum (A-C) interface is a well-known obstacle that impedes the development of such methods for dynamic problems [54]. As mentioned above, CAC is a dynamic multiscale method with a unified formulation of balance laws that govern both atomistic and continuum regions [17-19]. Therefore, it reduces the phonon wave reflection problem at the A-C interface to a numerical problem caused by different FE mesh. The phonon dispersion relation in Fig.1 indicates that CAC is able to simulate accurately the dynamics of phonons whose wavelength is larger than a critical wavelength $2\pi/k_c$.

To further verify this capability of the CAC method, the CAC simulation results of the propagation of a heat pulse in a single crystal model that has a concurrent atomic-coarse mesh, i.e., non-uniform mesh, are compared with the one with a uniform FE coarse mesh. The heat pulse is constructed to continuously generate phonons with wavelengths ranging from 5nm to 250nm and is applied to the center of the models where the coarse elements are employed, with the spot size of the heater being 10nm to avoid strong phonon-phonon scattering. To eliminate wave reflections from the boundaries of the computer models, phonon waves that reach the specimen boundaries are absorbed by applying damping at regions near the free surfaces of the specimens. In this case, a cubic polynomial damping function with gradual changing viscosity [55] is chosen. This is then equivalently to measure the transient propagation of heat pulse in a part of a larger specimen. Note that in order to monitor heat flow in specimens for a long time duration, this technique is used in all the simulations presented in this work.

In Fig. 5 we present the time sequence of the normalized kinetic energy obtained from CAC simulation of transient phonon propagation in the single crystal model with a uniform mesh. The longitudinal and transverse wavefronts are indicated in Fig. 5(c). The associated wave speeds are computed as ~6700 m/s and ~3900 m/s, respectively, consistent with the phonon group velocities of the LA and TA phonons near the Γ-point. The same process is also simulated in the single crystal model with a non-uniform mesh. In Fig. 6 we compare the normalized kinetic energy distribution in the two different computer models at a same time step, and find there is no noticeable difference. The maximum error is then quantified by comparing the total kinetic energy in different regions, and is less than 0.5%. This result confirms that the atomistic-continuum interface in CAC does not provide a numerical barrier for the dynamic propagation of long wavelength phonons. It is also noticed that relative short wavelength phonons channel along the [112] crystal directions, resulting in phonon-focusing "caustics", which are directions with especially high phonon fluxes. This indicates that CAC method is able to capture the experimentally observed phonon focusing phenomenon [50]. Note that the classical definition of temperature is made based on an equilibrium distribution of heat carriers, in this work, with the presence of transient ballistic phonon transport, the normalized kinetic energy, or nondimensional kinetic energy, is instead used to present the nonequilibrium process. Here the ballistic transport can be defined as the transport without scattering before it reaches the boundary [56].

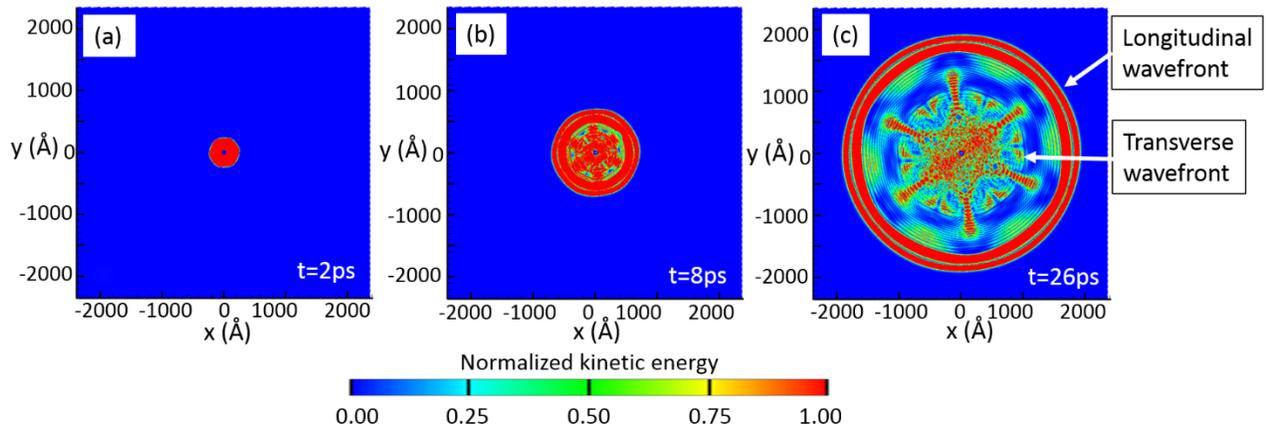

**Fig. 5** Propagation of continuous waves in the single crystal specimen. Contours are the normalized kinetic energy distribution. The longitudinal and transverse wavefronts are indicated in (c).

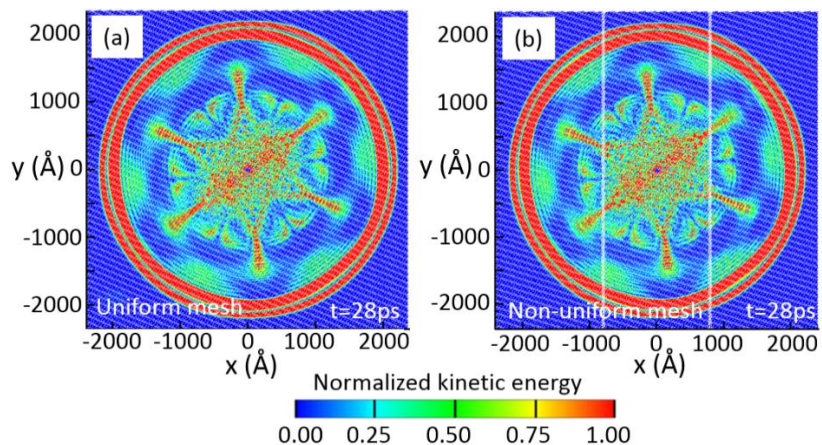

**Fig. 6** Comparison of the phonon wave propagation in the single crystal specimen constructed with (a) a uniform coarse mesh and (b) a non-uniform mesh with both atomic and coarse-scale finite element mesh.

## 3. Simulation of Phonon Pulse Propagation in Single Crystals and across GBs

### 3.1 Phonon pulse – grain boundary interaction

To study phonon-GB interaction during the propagation of a heat pulse, a heat pulse is constructed and applied to the heating region with a spot size of a 100nm for 2ps. Different from the continuous wave excitation in the simulations presented in Section 2, the phonon pulse here is applied to a large spot for a short time duration. This permits phonons to scatter with each other before the heat pulse propagating out of the heat source region, resulting in a large portion of diffusive phonons. In Fig. 7 we present the time sequence of the normalized kinetic energy obtained from CAC simulations of transient heat flow in the three material samples, the single crystal, GB-1 and GB-2 models. The left column of Fig. 7 presents the time sequences of the kinetic energy evolution in the single crystalline model. The circular wavefronts provide the evidence that heat transport in the single crystal model is partially carried by coherent phonons propagating away from the heat source. Such typical coherent phonon wave propagation is considered to be the signature of ballistic heat transfer in crystalline solids. It is seen from the kinetic energy contour in the left column of Fig. 7 that phonons with relatively shorter wavelengths, or equivalently, lower group velocities, scatter with each other, resulting in diffusive thermal transport. The presence of simultaneous ballistic and diffusive thermal transport in crystalline materials under heat pulse is a result of the finite relaxation time of phonons. The CAC simulation result is consistent with a recent experimental observation [57] and a theoretical work [14].

The middle and right column of Fig. 7 respectively present the time evolution of the kinetic energy obtained from CAC simulations of the GB-1 and GB-2 models. With the presence of GBs, the material responds to the heat pulse in a different manner. Before the phonons reach the GBs, from t = 0 to t = 11 ps, neither kinetic energy discontinuities nor distorted wave-fronts are observed. At this stage, phonons in the GB models propagate in the same fashion as that in the single crystal model. After t = 11 ps, the fast propagating phonons reach the GBs and subsequently some phonons scatter by the GB while some transmit across the GB without scattering, as shown in Fig. 7(h-j) in the middle column and Fig. 7(m-o) in the right column, indicating phonon transport across GBs is also partially coherent and partially diffusive. Note that the near-circular-shaped energy "crests" of the ballistic phonons transmit across the GBs coherently, but the energy "caustics" are deflected by the first GB into another direction after transmission across the GB. The new direction that the energy concentrates along is still [112] crystallographic direction, but in a different grain. The angle between the focusing directions in the adjacent crystalline grains is consistent with the inclination angle of the two grains, which is 13.2°. Additionally, some of the diffusively propagating phonons are trapped within the grain at the center of the specimen. The combination of the diffusive scattering of coherent and incoherent phonons and the reflection of the coherent phonons by the GBs is of strongly nonequilibrium nature, and leads to the Kapitza resistance to heat flow. This is manifested as the discontinuities of the kinetic energy at the GBs, which can be clearly observed in Fig. 7(h-j) in the middle column and Fig. 7(m-o) in the right column. At the end of the simulation, when further heat flow becomes negligible, the thermal transport in the specimens with GBs is nearly diffusive in character, with the discontinuity in kinetic energy along the 1st GB being much more significant than that along the 2nd GB.

To gain a quantitative understanding of the GB resistance to heat flow, In Fig. 8 we plot the time history of the accumulated percentage of transmitted kinetic energy, i.e., the kinetic energy that has been transmitted across the GBs divided by the total kinetic energy initially excited by the heat pulse. The transmitted energy is obtained through calculation of the energy that has not been transmitted across the GBs, i.e., the total kinetic energy initially excited by the heat pulse $E_0$, minus the kinetic energy within the yellow dash-line box region marked on the GB-2 model in Fig. 2(b), $E_2(t)$. This gives the percentage of transmitted energy to be: $1 - E_2(t)/E_0$. Although there is no GB in the single crystal model and the GB-1 model does not have a second grain, for the purpose of comparison, the total kinetic energy within the same region of the single crystal and that of GB-1 models are also monitored. It is seen from Fig. 8 that, before t = 21 ps, no phonon has yet reached the 2nd GB, and the percentage of the transmitted energy is zero. From t = 21 ps to t = 33 ps, most of the fast propagating long-wavelength phonons transmit across the GBs coherently, correspondingly, there is no obvious difference between the energy transmission in the three models. It indicates that the GBs in both models are almost transparent to the long-wavelength phonons. At t = 33 ps, the energy caustics reach the GBs, and the GBs start to scatter phonons; as a result, the energy transmission curves start to differ from each other. The phonon transport across GBs in this time period is partially coherent and partially incoherent. The co-existence of coherent and incoherent phonon transport is observed until around t = 110 ps in the models with GBs, and we estimate that the contribution of coherent phonons to the transmitted energy is about 40%. At the end of the simulation, at t = 225 ps, when there is no further appreciable heat transfer, the transmitted energy is 61.3% and 51.5% in the GB-1 and GB-2 model, respectively. Therefore, the presence of the GBs has significantly changed the nature of phonon transport, with the long-wavelength coherent phonons dominating the heat transfer. To quantify the GBs contribution to the thermal resistance, we denote the ratio of the transmitted energy in the GB-1 and GB-2 models to that in the single crystal model as r, and then calculate the reduction of the energy transmission due to the presence of the GBs as $1 - r$. At the end of the simulations, they are respectively 32.4% and 43.2% for the GB-1 and GB-2 models. Thus, compared to single crystal, with a presence of one GB on each side, there is 32.4% reduction in energy transmission, and with two GBs, the reduction in energy transmission is about 43.2%.

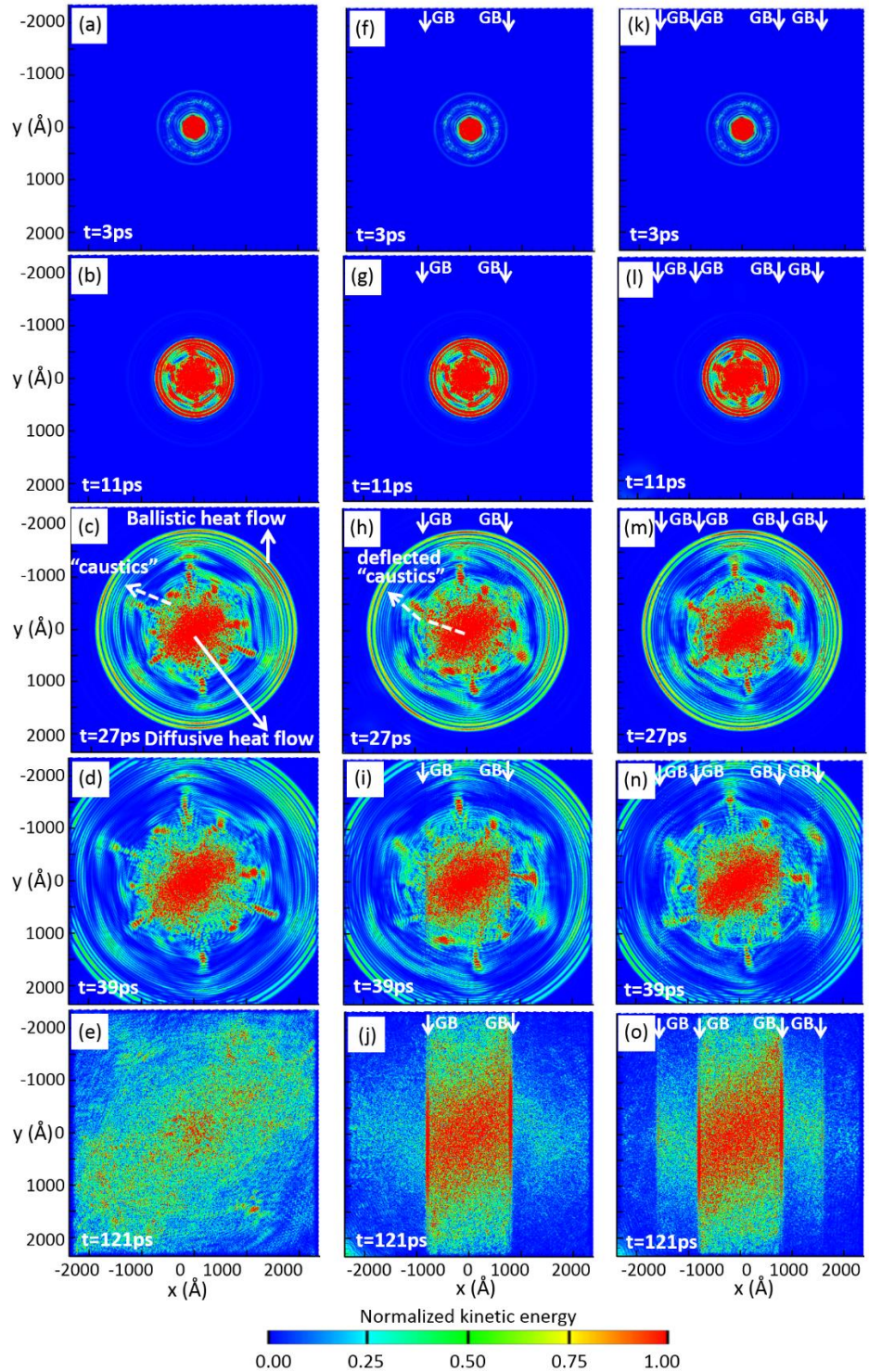

**Fig. 7** Time sequences of the normalized kinetic energy obtained from CAC simulations of transient heat flow in single crystal (left column), GB-1 (middle column) and GB-2 (right column) models (the location of GBs are indicated by white arrows), showing simultaneous ballistic and diffusive thermal transport, as well as the energy caustics as a result of phonon focusing (indicated by the white arrows in (c)), and the caustics are deflected by the GBs (indicated by the dashed white arrow in (h)).

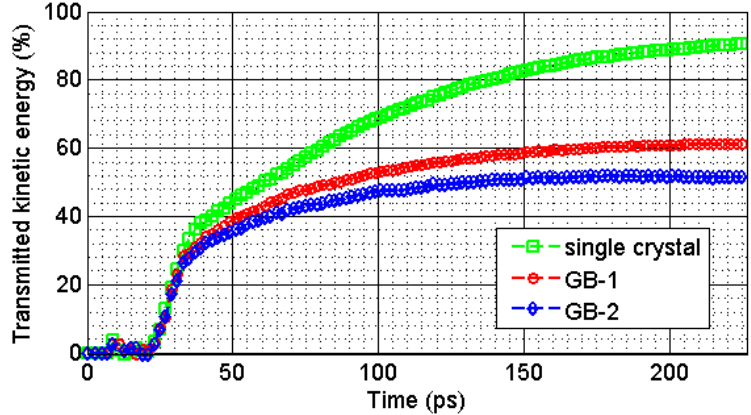

**Fig. 8** Time history of the accumulated percentage of transmitted kinetic energy.

### 3.2 Wavelength dependent energy transmission

The simulation results presented in the previous section indicate that the long-wavelength, low-frequency phonons transmit across GBs almost coherently, whereas the short-wavelength, high-frequency phonons are confined by the GBs. This wavelength-dependent wave transmission across GBs can be further quantified through continuous monochromatic wave simulations. For this purpose, we simulate the propagation of three continuous monochromatic waves with wavelength $\lambda$=20nm, 10nm and 5nm, respectively, in the GB-1 model. For comparison, the same three monochromatic wave simulations are also conducted on the single crystal model.

In Fig. 9 we present the time evolution of normalized kinetic energy for the characterization of the effect of GBs on phonon transport. The normalized kinetic energy is calculated as the total kinetic energy in the heat-source-grain divided by the maximum kinetic energy, or energy plateau that can be attained in the same region in the single crystal model $KE_{max}$. The $KE_{max}$ is reached in the single crystal when the continuous kinetic energy input in the "heat-source-grain" region is equal to the continuous energy flowing out. The energy flows in that region then reaches a steady-state, manifested as the plateau of the blue curves in Fig.8. The snapshots of kinetic energy distribution in GB-1 model at the end of each simulation are presented in Fig. 10(a-c). It is seen from Fig. 10(a) that the 20nm-phonon can transmit across the GBs almost transparently, and the kinetic energy evolution curve in Fig. 9(a) is almost identical to that of the single crystal model. The 10nm-phonons shown in Fig. 10(b) transmits across the GB with a small portion of waves reflected back into the heat-source-grain, giving rise to ~5% of difference in the energy evolution curves in Fig. 9(b). When the wavelength reduces to 5nm, as can be seen from Fig. 10(c), a large amount of energy is reflected back and trapped within the heat-source-grain, resulting in ~50% difference in kinetic energy at the end of the simulation as shown in Fig. 9(c). Therefore, the monochromatic wave simulations in this section confirm that the GBs confine short-wavelength phonons with their wavelengths below around 10nm. We notice that 10nm is very close to the distance between adjacent pentagonal repeat units in the GB, which is 11nm. A future study on different GBs with different density of repeat units will be able to reveal more information.

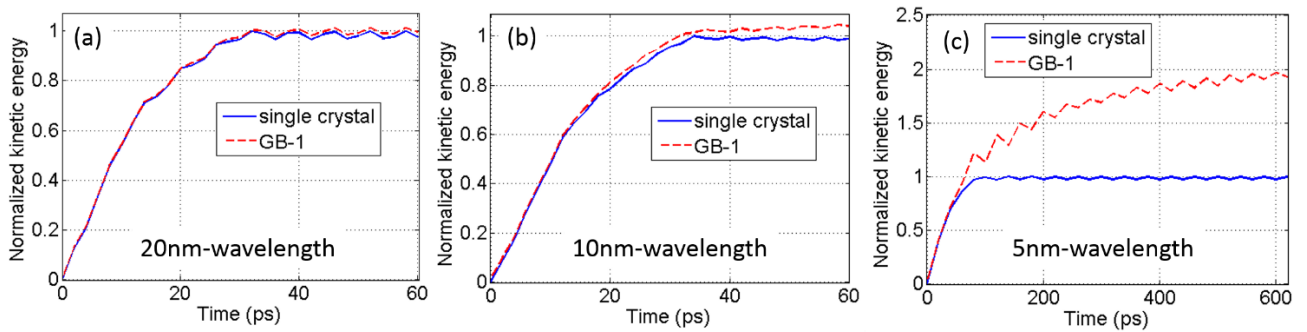

**Fig. 9** Comparison of the kinetic energy evolution (normalized by the value of the energy plateau in the single crystal model) in heat-source-grain of the GB-1 model and the same region in the single crystal model. (a-c) are the results of three monochromatic wave simulations with their wavelength at (a) 20nm, (b) 10nm and (c) 5nm.

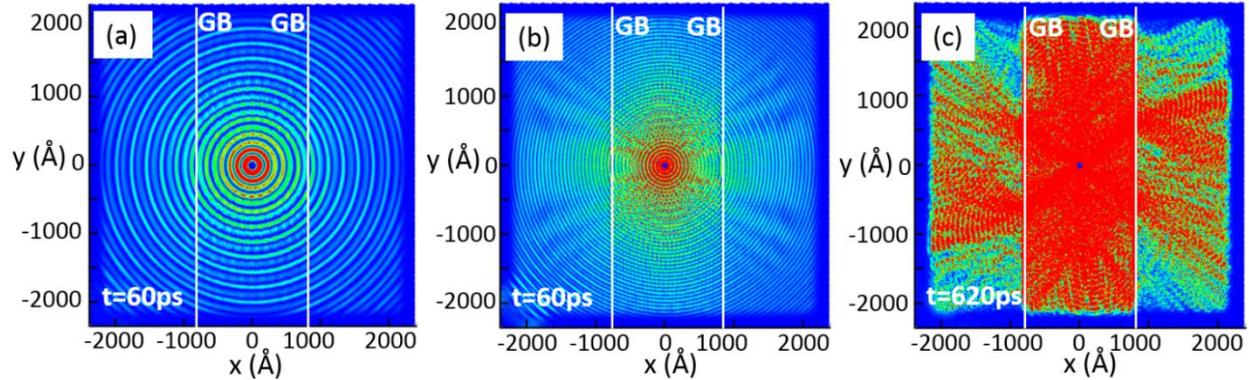

**Fig. 10** Normalized kinetic energy distribution in the GB-1 model for three different monochromatic wave propagation, their wavelengths are respectively (a) 20nm, (b) 10nm and (c) 5nm. The snapshots are taken at the end of each simulation.

### 3.3 Phonon-induced GB structural change

In Fig. 11, we present snapshots of the time sequence of the motion of atoms obtained through a mapping from the mesh deformation during the CAC simulations. Results show that the phonon-GB interaction gives rise to the GB structural change, depending on the intensity of the heat pulse. With a low intensity, the GB structure is not affected; while with an increased intensity of heat pulse, the GB is found to reconstruct locally, but recovers after the wavefront passes by; further increasing the intensity of heat pulse, some local regions of the GB are permanently reconstructed, resulting in a wavy GB after the phonon-GB interaction. In Fig. 11(a)-(c), we present the GB structural change as a result of its interaction with the phonons. In Fig. 11(a), we see a straight GB before the GB interaction with the phonon pulse, while in Fig. 11(b), the GB is reconstructing during the process of GB interaction with the fast phonons. After the transmission of the fast phonons across the GB, it can be seen from Fig. 11(c) that the GB structure has been recovered. An analysis of the location where the GB reconstructs reveals that it is the "caustics" possessing high energy flux that forces the GB to reconstruct, while the near-circular shaped energy crests transmit across the GB almost transparently, leaving the GB structure unchanged.

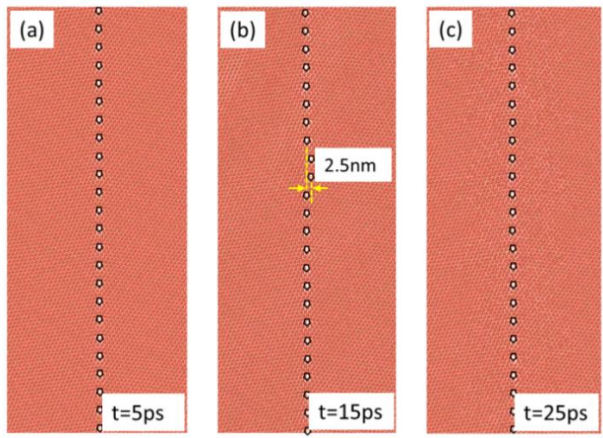

**Fig. 11** Snapshots of the time sequences of the atomic arrangements showing the GB structural change due to the phonon-GBs scattering.

### 4. Summary

In summary, we have presented a coarse-grained atomistic study of transient phonon thermal transport in materials with grain boundaries under ultrafast heat pulse excitation. In addition to the demonstration of a new computational method for phonon transport, this work has provided both qualitative and quantitative results of phonon-GB interaction during the prop-agation of a heat pulse, reproduced several important physical phenomena that have never been captured in a simulation, which facilitates the understanding of the physics of phonon transport. Major findings are summarized as follows:

With a heat pulse that is constructed based on a Bose-Einstein distribution of phonons and contains phonons of wavelength ranging from 5 nm to 250nm, we have reproduced the remarkable phenomenon of phonon focusing, which is the tendency for the ballistic heat flux emitted from a point source to concentrate along certain directions of the crystal [50]. Phonon focusing is a transient phonon transport phenomenon, quite different from the steady-state thermal transport. While steady-state thermal transport may still obey the Fourier's heat conduction law, this work shows that the transient phonon transport is a highly nonequilibrium process; a constant temperature gradient does not exist in transient phonon transport, and hence the Fourier heat conduction law no longer applies.

The spatially and temporally resolved simulation results provide a visualization of the transient dynamic process of the propagation of a heat pulse and that of the phonon-GB interaction. The results of three different simulated specimens have demonstrated that the propagation of a heat pulse in a crystalline solid with or without grain boundaries is partially ballistic and partially diffusive, i.e., there is a co-existence of ballistic and diffusive thermal transport, with the long-wavelength phonons traveling ballistically while the short-wavelength phonons scatter with each other and travel diffusively. When meeting with a GB, long wavelength phonons transmit across the GB coherently, most of the focused phonons in the energy caustics are deflected by the GBs, and short wavelength phonons are reflected back by the GBs and confined within the heat-source-grain.

The phonon-GB interactions give rise to the discontinuities in the kinetic energy at GBs, which provide a clear evidence of the GB thermal boundary resistance. The ratio of the energy that has transmitted across the GBs with respect to the total kinetic energy excited as well as that with respect to energy transmission in the single crystal has been quantified, thus providing a quantitative understanding of the GB resistance to phonon thermal transport. The quantitative measurement, together with the visualization reveals it is the long-wavelength phonons propagating coherently in the specimen that act as the major heat carriers.

The simulations in this work have also demonstrated that the GB interaction with phonons leads to a local reconstruction of the GBs, while may or may not be recovered, depending on the intensity of the phonon pulse. This structure change is driven by the focused energy caustics of the phonons, mainly related to long-wavelength phonons. It is observed that the structure of the GB can be recovered after partial transmission of the energy caustics. The high intensity of a heat pulse, however, can lead to a permanent GB reconstruction. The role of long-wavelength phonons on the dynamic behavior of the GBs can thus be significant.

To the best of our knowledge, this work is the first attempt using a coarse-grained atomistic method to study and visualize phonon transport across GBs. This study shows that a coarse-grained simulation can still capture the essential physics of phonon dynamics and phonon transport, including phonon focusing, co-existence of ballistic and diffusive phonon transport, non-Fourier heat conduction in transient transport, GB-phonon interaction, and local reconstruction of the GBs, as a result of the long-wavelength phonon-GB interaction. If the short-wavelength phonons are included, we expect that the quantitative energy transmission coefficients would be affected, but the qualitative results shall remain, because the shorter wavelength phonons shall also mostly be confined by the first interface, scatter with each other and travel diffusively. Therefore, the coarse-grained atomistic simulation results presented in this work provide a full qualitative picture of the heat pulse propagating across GBs, as well as producing quantitative results complementary to fully atomistic simulations at the nanoscale.

**AUTHOR INFORMATION**

**Corresponding Author**
* E-mail: xiangchen@ufl.edu

**Author Contributions**
All authors have given approval to the final version of the manuscript.

**Notes**
The authors declare no competing financial interest.

## ACKNOWLEDGMENT

This material is based upon research supported by the U.S. Department of Energy, Office of Science, Basic Energy Sciences, Division of Materials Sciences and Engineering under Award # DE-SC0006539.